\documentclass[sn-mathphys,Numbered]{sn-jnl}
\usepackage{graphicx}%
\usepackage{enumitem}
\newlist{legal}{enumerate}{10}
\setlist[legal]{label*= \arabic*.}
\usepackage[export]{adjustbox}
\usepackage{multirow}%
\usepackage{amsmath,amssymb,amsfonts}%
\usepackage{mathtools}
\usepackage{amsthm}%
\usepackage{mathrsfs}%
\usepackage[title]{appendix}%
\usepackage{xcolor}%
\usepackage{cancel}
\usepackage{textcomp}%
\usepackage{manyfoot}%
\usepackage{booktabs}%
\usepackage{algorithm}%
\usepackage{algorithmicx}%
\usepackage{algpseudocode}%
\usepackage{listings}%
\usepackage{musicography}
\usepackage{ulem}
\theoremstyle{thmstyleone}%
\begin{document}

\title[Article Title]{Regular black hole from regular initial data}


\author[]{\fnm{Karim} \sur{Mosani$^1$}}%\email{kmosani2014@gmail.com}

\affil[]{\orgdiv{$^1$Department of Mathematics}, \orgname{University of Tuebingen}, \orgaddress{\street{Auf der Morgenstelle 10}, \city{Tuebingen}, \postcode{72076},  \country{Germany}}}

\author[]{\fnm{Pankaj S} \sur{Joshi$^2$}}%\email{kmosani2014@gmail.com}
\affil[]{\orgdiv{$^2$International Centre for Space and Cosmology, School of Arts and Sciences}, \orgname{Ahmedabad University}, \orgaddress{\street{} \city{Ahmedabad}, \postcode{380009},  \country{India}}}


\abstract{Recently, there has been an interest in exploring black holes that are regular in the sense that the central curvature singularity is avoided. Here, we depict a method to obtain a regular black hole (RBH) spacetime from the unhindered gravitational collapse, beginning with regular initial data of a spherically symmetric perfect fluid. In other words, we obtain the equilibrium (static)  spacetime $(\mathcal{M}, \Tilde{g})$ as a limiting case of the time-evolving (non-stationary) spacetime $(\mathcal{M}, g)$. In the spirit of Joshi, Malafarina and Narayan (\textit{Class. Quantum Grav. 31, 015002, 2014}), our description of gravitational collapse is implicit in nature in the sense that we do not describe the data at each time-slice. Rather, we impose a condition in terms of geometric and matter variables for the collapse to have an end-state that is devoid of incomplete geodesics but admits a marginally trapped surface (MTS). The admission of MTS causally disconnects two mutually exclusive regions $\Hat{\mathcal{M}}_1$ and $\Hat{\mathcal{M}}_2\subset \mathcal{M}$ in the sense that $\forall~p\in\Hat{\mathcal{M}_2}$, the causal past of $p$ does not intersect $\Hat{\mathcal{M}}_1$.  While the classic Oppenheimer-Snyder collapse model necessarily produces a black hole with a Schwarzschild singularity at the centre, we show here that there are classes of regular initial conditions for which the collapse gives rise to a RBH. }

%\keywords{Black Hole, Stationary Spacetime, Axisymmetric Spacetime, Trapped Photon Region}



\maketitle
%\tableofcontents
\section{Introduction}
The Penrose and Hawking incompleteness theorems
\cite{Hawking_1973}
prove the existence of incomplete causal geodesics $\gamma\subset \mathcal{M}$ (where $\mathcal{M}$ is a real smooth manifold) under certain conditions. The general pattern of the theorems (that we call here the pattern incompleteness theorem) is as follows
\cite{Senovilla_1998}: If $(\mathcal{M},g)$ (four-dimensional spacetime) satisfies
1) a causality condition,
2) a curvature condition, and
3) a suitable initial/boundary condition,
then $\exists$ incomplete causal geodesics. One of the incompleteness theorem states
\cite{Hawking_1973}: If $(\mathcal{M},g)$ is 1) connected, globally hyperbolic spacetime with non-compact Cauchy surface $\Sigma$, 2) $Ric\left(\zeta,\zeta\right)\geq 0$ $\forall$ $\zeta$ ($Ric$: Ricci tensor, $\zeta$: null vector), and 3) admits a trapped surface $T$, then at least one inextendible future-directed orthogonal null geodesic from $T$ exists that is future incomplete.

One interpretation of these theorems is that under certain scenarios, general relativity predicts its own breakdown, and hence, one should seek a way of getting rid of geodesic incompleteness. 
%\sout{It should be noted that in this paper, we do not promote or demote this interpretation. However,} 
If we proceed with this interpretation, we must consider only those initial data that have the following property: At least one of the three conditions of the pattern incompleteness theorem should not be satisfied for the maximal Cauchy development $(\mathcal{M},g)$ of $\Sigma$. 
%We must necessarily disobey at least one of the three conditions imposed on the \textcolor{black}{maximal Cauchy development} $(\mathcal{M},g)$ of the initial data, mentioned in the pattern incompleteness theorem. 
This restriction on the initial data leads to avoidance of the existence of incomplete causal geodesics. In this paper, we say that a spacetime admits a \textit{singularity} if it is geodesically incomplete and the curvature scalars diverge along any incomplete geodesic $\gamma$ as one approaches the limit point of $\gamma$ that $\not \in \mathcal{M}$.

In the astrophysics/cosmology sector, the formation of black holes in the early universe have been studied in \cite{Khlopov1, Khlopov2, Khlopov3}. Additionally, recent observations of the central compact object of the M87 galaxy
\cite{EHT_1}
and that of our own galaxy
\cite{EHT_6}, 
earlier observations, such as that of Sagittarius A* at our galactic centre
\cite{SGR}, and similar observations suggest an extremely strong gravity region at the galactic centre. Such strong gravity region could admit a singularity. However, if we proceed with the interpretation of the incompleteness theorems mentioned above and wish to avoid the breakdown of general relativity, then at the galactic centre, one should have an ``extremely dense" object having astrophysical signatures similar to a black hole, but the spacetime being geodesically complete, let alone admitting a singularity.

Attempts have been made in the past regarding ways to avoid geodesic incompleteness in black hole solutions by the conformal transformation of the original black hole metric in a certain way
\cite{Bambi_2017, Chakrabarty_2018}. 
These are objects with a horizon but complete. Such objects are dubbed regular black holes. Other regularization techniques have obtained a singularity-free analogue of singular spacetime metrics with a horizon (see \cite{BambiRBH, Rubio_2025} and the references therein for a recent summary of the present understanding on the subject). It is worth noting that the author in \cite{Waliachemestry} explores the concept of polymerized black holes inspired by loop quantum gravity. This approach aims to address the singularities inherent in classical black hole models by introducing quantum corrections that result in RBH.

Proposed RBH model spacetimes have also been studied from an observational point of view in the context of shadows \cite{Walia1, Walia2, Walia3}, gravitational lensing \cite{Walia4}, and photon ring structure \cite{Walia5}. It is worth noting that not all these models are geodesically complete. For e.g. Hayward 
\cite{Hayward}, 
and Culetu-Simpson-Visser \cite{Culetu1, Culetu2, Simpson} spacetimes 
are shown to be geodesically incomplete by Zhou and Modesto
\cite{Modesto_2022}.

Gravitational collapse is the key astrophysical mechanism that generates stars, galaxies, and other structures in the universe. However, there have been few attempts to understand the formation of RBHs as an end state of gravitational collapse. Its dynamical formation using the quantum gravity approach has been discussed in \cite{Cassadio, Torres, Kelly, Hussain, Ashtekar}. In \cite{Malafarina_2022sca}, the author took a semi-classical approach to resolving singularities and obtained a RBH as an end state of dust collapse. No attempts have been made so far (as of the completion of this article) to understand the formation of RBH within the framework of general relativity.

In this article, we show a method of obtaining a RBH as an end-state of gravitational collapse from regular initial data in general relativity. By regular initial data, we mean that we have a well-defined, suitably differentiable initial profiles on the three-dimensional spacelike hypersurface $\Sigma \subset \mathcal{M}$ that foliates $\mathcal{M}$ ($\mathcal{M}=\Sigma \times \mathbb{R}$). Here, we employ the method of obtaining an equilibrium configuration previously studied in \cite{JMN2} 
and put the following two restrictions so that this end-state equilibrium is a RBH: 1) The equilibrium configuration should be regular, i.e. it should not admit incomplete geodesics (let alone a curvature singularity), and 2) It should have a horizon, namely a marginally trapped surface (MTS) so that no future-directed null geodesics should exist from a point enclosed by the MTS to a point that is not enclosed by the MTS. To satisfy the first restriction, we will compensate with condition 3) of the pattern singularity theorem. This initial/boundary condition demands the existence of trapped surfaces.  

%\textcolor{black}{The sections are organized as follows: $II$) We formulate the definition of equilibrium configuration, and in terms of the proposed definition, we define a RBH spacetime. $III$) We begin with a spacetime $(\mathcal{M},g)$ corresponding to perfect fluid and admitting spherical symmetry, and construct an equilibrium configuration $(\mathcal{M},\Tilde{g})$. $IV$) We exploit the freedoms in Einstein's equation and set restrictions such that $(\mathcal{M},\Tilde{g})$ admits a RBH; we also discuss its properties. $V)$ We conclude.}

%\textcolor{black}{We aim to construct a spacetime $(\mathcal{M},g)$ that models a spherically symmetric cloud that undergoes gravitational collapse. We want this cloud to be bounded in $\Sigma$.}
The phenomena of gravitational collapse is modelled %by constructing a spacetime $(M,g)$ 
using two spacetimes $(\mathcal{M}_1,g_1)$ (interior) and $(\mathcal{M}_2,g_2)$ (exterior) with oriented boundaries $\partial \mathcal{M}_1$ and $\partial \mathcal{M}_2$ respectively such that $\partial \mathcal{M}_1$ is diffeomorphic to $\partial \mathcal{M}_2$, i.e. $\partial \mathcal{M}_1\cong \partial \mathcal{M}_2\cong \Xi$. A new spacetime $(\mathcal{M},g)$ is then constructed such that $\mathcal{M}$ is a disjoint union of $\mathcal{M}_1$ and $\mathcal{M}_2$. $\partial \mathcal{M}_1$ and $\partial \mathcal{M}_2$ are embedded in $\mathcal{M}_1$ and $\mathcal{M}_2$, respectively, and their points are identified such that Israel-Darmois junction conditions
\cite{Darmios_27, Israel_67}
are satisfied 
\cite{Mena_11}. Let ${^{\Xi}}{g}{_1}$ denote the metric induced on $\Xi \hookrightarrow \mathcal{M}_1$ and ${^{\Xi}}{g}{_2}$ denote the metric induced on $\Xi \hookrightarrow \mathcal{M}_2$. Let 
$$
K_1: T\Xi \times T\Xi \to ~C^{\infty}\left(\Xi\right):~(X,Y)\mapsto g_1\left(\nabla_X \mathcal{V}, Y\right),
$$
and
$$
K_2: T\Xi \times T\Xi \to ~C^{\infty}\left(\Xi\right):~(X,Y)\mapsto g_2\left(\nabla_X \mathcal{V}, Y\right),
$$
be the extrinsic curvatures of $\Xi \hookrightarrow \mathcal{M}_1$ and $\Xi \hookrightarrow \mathcal{M}_2$ respectively (here $\mathcal{V}\in T\mathcal{M}$ is the global unit normal vector field of $\Xi$). The Israel-Darmois junction conditions are as follows:
\begin{align*}
    {^\Xi}{g}{_1}\equiv & {^\Xi}{g}{_2}, \\
    K_1\equiv & K_2.
\end{align*}

In section (II), we describe the gravitational collapse of interior perfect fluid glued to the exterior generalized Vaidya spacetime \cite{Wang_1999}  such that the abovementioned Israel-Darmois conditions are satisfied for the glueing timelike hypersurface. In section (III), we impose conditions in terms of geometric variable $R$ (to be defined later) to obtain an equilibrium spacetime. We further impose sufficient conditions on the matter variable $F$ (to be defined later) and specify the matching timelike hypersurface so that we obtain a RBH. We conclude the paper in section (IV). Here, we use the geometrized units $(c=8\pi \mathbb{G}=1)$. Notations for the spacetime metric: $g_1$, $g_2$, $g$, $\Tilde{g}_1$, $\Tilde{g}_2$, and $\Tilde{g}$ stands for the spacetime metrics corresponding to the interior collapsing spacetime, the exterior collapsing spacetime, the total collapsing spacetime, the interior equilibrium (RBH) spacetime, the exterior equilibrium (RBH) spacetime, and the total equilibrium spacetime, respectively.

\section{Dynamical spacetime} 
We model the gravitational collapse with the interior spacetime corresponding to spherically symmetric perfect fluid. To begin with, let us consider a spherical gravitational collapse of a general perfect fluid modelled by $(\mathcal{M}_1,g_1)$. The spacetime metric $g_1$ in comoving coordinate basis is described as
\begin{equation}\label{st1}
    ds^2=-e^{2\nu}dt^2+\frac{R'^2}{G}dr^2+R^2d\Omega^2.
\end{equation}
Here, $\nu=\nu(t,r)$, $R=R(t,r)$ and $G=G(t,r)$. The superscripts dot and prime above the functions in this context denote the partial derivative with respect to comoving coordinates $t$ and $r$, respectively. Additionally, we set $r_b\in\mathbb{R}^+$ such that (\ref{st1}) is valid for $0\leq r\leq r_b$.  Einstein's field equations for a perfect fluid are
\begin{equation}\label{efe1}
 \rho =\frac{F'}{R^2R'},
\end{equation}
\begin{equation}\label{efe2}
      p =-\frac{\dot F}{R^2\dot R},
\end{equation}
\begin{equation}\label{efe3}
    \nu '=-\frac{p'}{\rho +p},
\end{equation}
and
\begin{equation}\label{efe4}
    \dot G=2\frac{\nu'}{R'}\dot R G,
\end{equation}
where
\begin{equation}\label{msmf}
     F=R(1-G+e^{-2\nu}\dot R^2)
\end{equation}
is the Misner-Sharp mass function
\cite{Misner_1964, Szabados_2009}. The collapsing spherical cloud comprises concentric spherical shells, each identified by a comoving radial coordinate $r$. The physical interpretation of $F$ is that it gives us the notion of mass inside a spherical shell of radial coordinate $r$  at comoving time $t$. Further, we put the constraint $R'>0$ so that these shells do not 
%[*** Such abbreviations as `don't' not used in research papers, please use full forms, no exclamation marks etc]
cross each other. 

 It has been shown before that the Israel-Darmois junction condition is satisfied if the exterior spacetime is the generalized Vaidya spacetime
\cite{Wang_1999}
with a suitable generalized Vaidya mass function
\cite{Joshi_1999, Goswami_2007}. For the sake of completion, we mention the restriction on the generalized Vaidya mass imposed by the abovementioned condition. The exterior generalized Vaidya spacetime $(\mathcal{M}_2,g_2)$, expressed in retarded null coordinate $v$ is given by
\begin{equation}
    ds^2=-\left(1-\frac{2M_V(\mathcal{R},v)}{\mathcal{R}}\right)dv^2-2dvd\mathcal{R}+\mathcal{R}^2d\Omega^2.
\end{equation}
Here, $\mathcal{R}$ is called the generalized Vaidya radius. We define the boundary of $\mathcal{M}_1$ by $r=r_b>0$, and denote it by $\Xi$. We have $\Xi\hookrightarrow \mathcal{M}_1$ as well as $\Xi\hookrightarrow \mathcal{M}_2$ (upto diffeomorphism in the second case). We then have
\begin{equation}
    {^\Xi}{g}{_1}= -e^{2\nu}dt^2 +R^2d\Omega^2,
\end{equation}
and
\begin{equation}
     {^\Xi}{g}{_2}=-\left(1-\frac{2M_V(\mathcal{R},v)}{\mathcal{R}}+2\frac{d\mathcal{R}}{dv}\right)dv^2+\mathcal{R}^2d\Omega^2.
\end{equation}
Using the Israel-Darmois junction conditions mentioned above, we obtain
\begin{equation}\label{mc1}
    R(r_b,t)=\mathcal{R},
\end{equation}
\begin{equation}\label{mc2}
    F(t, r_b)=2M_V(\mathcal{R},v),
\end{equation}
\begin{equation}\label{mc3}
    \left(\frac{dv}{dt}\right)_{\Xi}=\frac{\sqrt{G} e^{\nu}+\dot R}{1-\frac{F}{R}},
\end{equation}
and 
\begin{equation}\label{gvm}
 M_V(\mathcal{R},v),_{\mathcal{R}}=\frac{F}{2R}+\frac{R\dot R e^{-2\nu}}{\sqrt{G}}+
\frac{R}{R'}e^{2\nu}\nu'\sqrt{G}.
\end{equation}
The last two equations are evaluated at $r=r_b$. The solution of Eq. (\ref{gvm}), after employing (\ref{mc1}-\ref{mc3}), describes $g_2$ completely. For a detailed explanation, refer to 
\cite{Goswami_2007}.

\section{Static equilibrium}
The collapsing cloud attends equilibrium configuration $\mathcal{E}=\left(\mathcal{M}_1,\Tilde{g}_1\right)$ if
\begin{equation}\label{equicond}
    \dot R=\ddot R=0.
\end{equation}
We denote the physical radius $R(t,r)$ of the cloud in $\mathcal{E}$ as $R_e(r)$  (subscript $e$ for the functions denotes their form in $\mathcal{E}$). We now have
\begin{equation}\label{tendto}
\begin{split}
    & R(t,r)\to R_e(r,), \hspace{0.4cm} F(t,r)\to F_e(r), \\  &\nu(t,r)\to \nu_e(r), \hspace{0.4cm} G(t,r)\to G_e(r).
\end{split}
\end{equation}
Imposing conditions (\ref{equicond}) and using Eq.(\ref{msmf}), we obtain the following two equations that fix the behaviour of metric coefficients in $\mathcal{E}$:
\begin{equation}
    G_e(r)=1-\frac{F_e(r)}{R_e(r)},
\end{equation}
and
\begin{equation}
    \left(G,_R\right)(r)=G,_R(r,R_e(r))=\frac{F_e}{R_e^2}-\frac{\left(F,_R\right)_e}{R_e}.
\end{equation}
In the coordinate basis of the newly defined coordinate system $(t,r) \mapsto (t,z)$, where $z=R_e(r)$, we write the functions $\mathcal{X}_e(r)$ in (\ref{tendto}) as $\Tilde{\mathcal{X}}(z)$. Einstein's equations (\ref{efe1}) and (\ref{efe3}) that $\mathcal{E}$ satisfies are
\begin{equation}\label{eq1}
    \Tilde{\rho}=\frac{\Tilde{F},_z}{z^2},
\end{equation}
and
\begin{equation}\label{eq3}
    \Tilde{p},_{z}=-(\Tilde{\rho}+\Tilde{p})\Tilde{\nu},_z.
\end{equation}
From Eq.(\ref{efe2}) and (\ref{efe4}), one obtains
\begin{equation}\label{eq2}
    \Tilde{p}=\frac{2\Tilde{\nu},_z}{z}\left(1-\frac{\Tilde{F}(z)}{z}\right)-\frac{\Tilde{F}(z)}{z^3},
\end{equation}
The spacetime metric $\Tilde{g}_1$ for $z\leq z_b=R_e(r_b)$ is given by
\cite{JMN2}
\begin{equation}\label{st2}
    ds^2=-e^{2\Tilde{\nu}(z)}dt^2+\frac{dz^2}{1-\frac{\Tilde{F}(z)}{z}}+z^2d\Omega^2.
\end{equation}
%\textcolor{black}{This matching is not smooth in the sense that the second Junction condition of Israel-Darmois is not satisfied. However, this violation comes with a sound physical interpretation that a surface layer with stress-energy tensor is present at the hypersurface
%\cite{Poisson}.}
We note that such $\mathcal{E}$ obtained as a result of the dynamical evolution of initial data is achieved in infinite comoving time because of the conditions that we begin with, i.e. Eq.(\ref{equicond})\footnote{Consider an example of a function $f(x)=a (x-b)^3$. Now, the first and second derivatives of $f$ with respect to $x$ are zero at $x=b$, but one does not achieve staticity for $x>b$ ($f$ continues evolving with $x$ for $x>b$). This example depicts that just by relying on the vanishing of the first two derivatives, one cannot expect to obtain the equilibrium configuration in the finite value of the domain.}.
%\textcolor{black}{Additionally, $r_b$ is that coordinate radius in the spacetime metric (\ref{st1})}
As the functions $\mathcal{X}(t,r)\to \mathcal{X}_e(r)=\tilde{\mathcal{X}}(z)$ in (\ref{tendto}), we have $g_1 \to \tilde{g}_1$.

One can eliminate $\Tilde{\nu}$ from (\ref{eq3}) and (\ref{eq2}), and obtain a first order differential equation of $\Tilde{p}(z)$ called Tolman-Oppenheimer-Volkoff
equation as
\begin{equation}\label{dep}
\Tilde{p},_{z}+\frac{\left(\Tilde{F}+\Tilde{p}z^3\right)\left(\Tilde{\rho}+\Tilde{p}\right)}{2z\left(z-\Tilde{F}\right)}=0.
\end{equation}
Choosing $\Tilde{F}(z)$ closes the system of equations describing $\mathcal{E}$. One can then solve the differential Eq.(\ref{dep}) for a suitable initial condition $\Tilde{p}(z_0)=\Tilde{p}_0$ to obtain $\Tilde{p}(z)$ that is the pressure profile (in terms of the newly defined radial coordinate $z$). Substituting the functional form of $\Tilde{F}(z)$ and $p(z)$ in the differential equation (\ref{eq2}), one can then obtain the remaining function $\nu(z)$ by solving it.
$\mathcal{E}$ identified by the functional form $\Tilde{F}(z)$ and the initial condition $\Tilde{p}(z_0)$ can be obtained as an end-state of a wide class of dynamics, each of which is further identified by one function $F(t,r)$ (or some other function, e.g. $p(t,r)$) such that $F(t,r)=F(r, R)\to F(r, R_e(r))=F_e(r)=\Tilde{F}(z)$ in the limit $t\to \infty$\footnote{Here we have committed a slight abuse of notation by using the same notation $F$ when we look at it as a function of $(t,r)$ as well as when we look at it as a function of $(r,R)$}.  
 
The necessary criterion for $\mathcal{E}$ to not admit a singularity is that the density $\Tilde{\rho}(z)$ should not blow up for any $z$. Typically, the density $\Tilde{\rho}(z)$ is monotonically non-increasing (stays the same or decreases) as we go away from the centre. 
If this is the case, then the boundedness of the density at the centre ensures its boundedness throughout the spacelike hypersurface identified by a constant $t$. 

From Eq.(\ref{eq1}) we have the following restriction on $\Tilde{F}(z)$:
\begin{equation}\label{orderF}
    \mathcal{O}(\Tilde{F})\sim \mathcal{O}(z^3).
\end{equation}
This restriction ensures that the density is bounded at the centre of the $\mathcal{E}$ (identified by $z=0$). In other words,
\begin{equation}
    \Tilde{F}(z)=\Tilde{F_0} z^3 +\Sigma_{i=1}^{\infty} \Tilde{F}_i z^{(i+3)}, \hspace{0.4cm} \Tilde{F_0}>0, \hspace{0.4cm} \Tilde{F}_i\in \mathbb{R},
\end{equation}
assuming $\Tilde{F}\in C^{\infty}(\mathcal{M}_1)$. Of course, $\Tilde{F}_i$'s are such that the equilibrium density $\Tilde{\rho}$ is a non-increasing function of $z$. Also, the pressure is smooth throughout the equilibrium matter field for a regular end-state. 
%The pressure profile can hence be expressed as Taylor expansion around $z=0$ as
%\begin{equation}
%    \Tilde{p}= \Sigma_{i=0}^{\infty}\Tilde{p}_i z^i.
%\end{equation}
%Solving the differential Eq.(\ref{dep}) gives us the coefficients $p_i$ as the recurrence relation
%\begin{equation}
%\begin{split}
%   & \Tilde{p}_{k+3}=\frac{1}{2\left(k+3\right)}\times 
%  \left(\Sigma_{l=0}^{k}\left(2\Tilde{F}_l\left(k-l+1\right)\Tilde{p}_{k-l+1}-\\
%&\left(\Tilde{F}_l\left(l+3\right)+\Tilde{p}_l\right)\left(\Tilde{p}_{k-l+1}+\Tilde{F}_{k-l+1}\right)\right)- \\ 
%&\left(\left(k+4\right)\Tilde{F}_{k+1}+ \Tilde{p}_{k+1}\right) \left(\Tilde{p}_0+ F_0\right)\right) \\
%& \forall \hspace{0.2cm} k\in \{0\}\cup \mathbb{N}.
%\end{split}
%\end{equation}
%Here $\Tilde{p}_0$ is free to choose, $\Tilde{p}_1=0$, and $\Tilde{p}_2$ is
%\begin{equation}
%    \Tilde{p}_2=-\frac{\left(\Tilde{p}_0+\Tilde{F}_0\right)\left(\Tilde{p}_0+3\Tilde{F}_0\right)}{4}.
%\end{equation}

For obtaining a RBH, one also requires the formation of at least one MTS, as mentioned before. We define such a surface in terms of the expansion of null geodesic congruence through this surface. This expansion  at a point $p\in \mathcal{O}\subset \mathcal{M}_1$ ($\mathcal{O}$ is open) is defined as the trace of the null Weingarten map 
$$\mathcal{W}: V_p~\to V_p~: ~\Bar{X}\mapsto \mathcal{W}(\Bar{X}):=\nabla_{\Bar{X}}\xi:=\overline{\nabla_{X}\xi}.$$
%purely spatial, rank two tensor field $\nabla_j \xi_i$
Here $\xi: \mathcal{M}_1 \supset \mathcal{O} \to T\mathcal{O}$ is a null vector field of tangents to the congruence, and $V_p$ is the two-dimensional fibre of the quotient set $T\mathcal{H}/\sim$ of the tangent bundle 
$$T\mathcal{H}:=\{Z\in T\mathcal{O}~\vert~\Tilde{g}_1~(Z,\xi)=0\},$$ 
with the equivalence relation $\sim$ defined as follows: 
$$
\forall ~X, Y\in T\mathcal{O}, ~X\sim Y \iff X-Y\propto \xi.$$
$\Bar{X}\in V_p$ is then the equivalence class of $X\in T\mathcal{H}$ corresponding to the above equivalence relation.
%%%%%%%%%%%%%%%%%%%%%%%%%%%%%%%%%%%%%%%%%%%%%%%%%%%%%%%%%%%%%%%%%%
\begin{figure}\label{omega}
  %  \centering
        \includegraphics[scale=0.45]{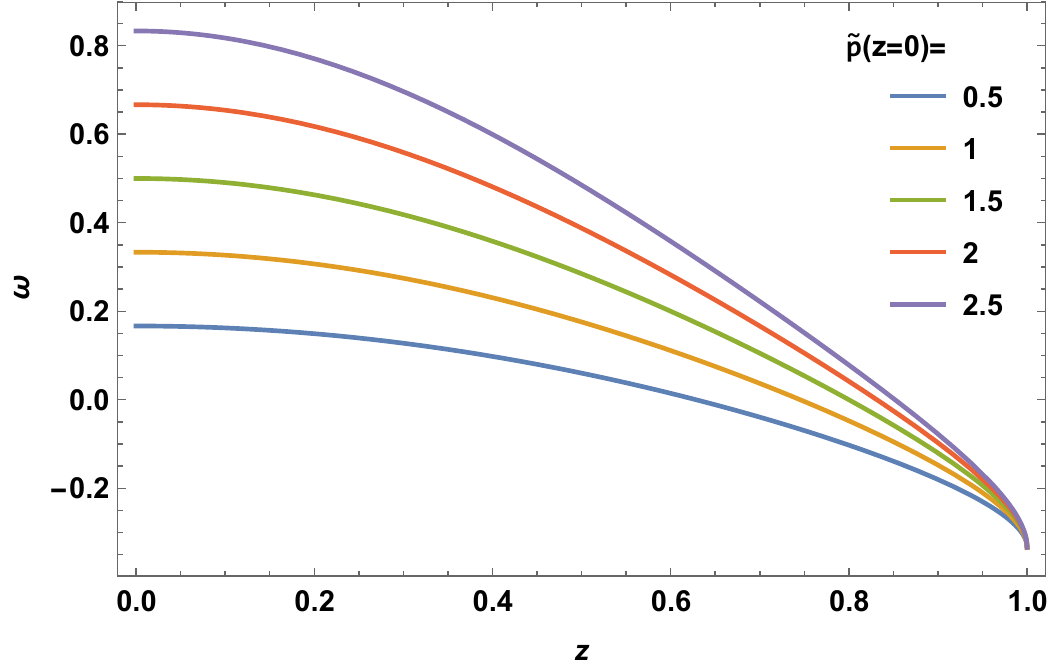}
        %\caption{Potential $A_V$ as a function of $a$.}
    \caption{Equation of state parameter ($\omega=\Tilde{p}/\Tilde{\rho}$) versus the newly defined radial coordinate $z$ corresponding to $\Tilde{F}(z)=\Tilde{F}_0z^3$, $\Tilde{F}_0=1$, and suitable initial condition for pressure $\Tilde{p}(z=0)$. For larger values of $z$, $\omega$ (and hence the pressure) becomes negative.}
    \label{rfidtag_testing}
\end{figure}
%%%%%%%%%%%%%%%%%%%%%%%%%%%%%%%%%%%%%%%%%%%%%%%%%%%%%%%%%%%%%%%%%%%%%%
\begin{figure}\label{ec}
  %  \centering
        \includegraphics[scale=0.4]{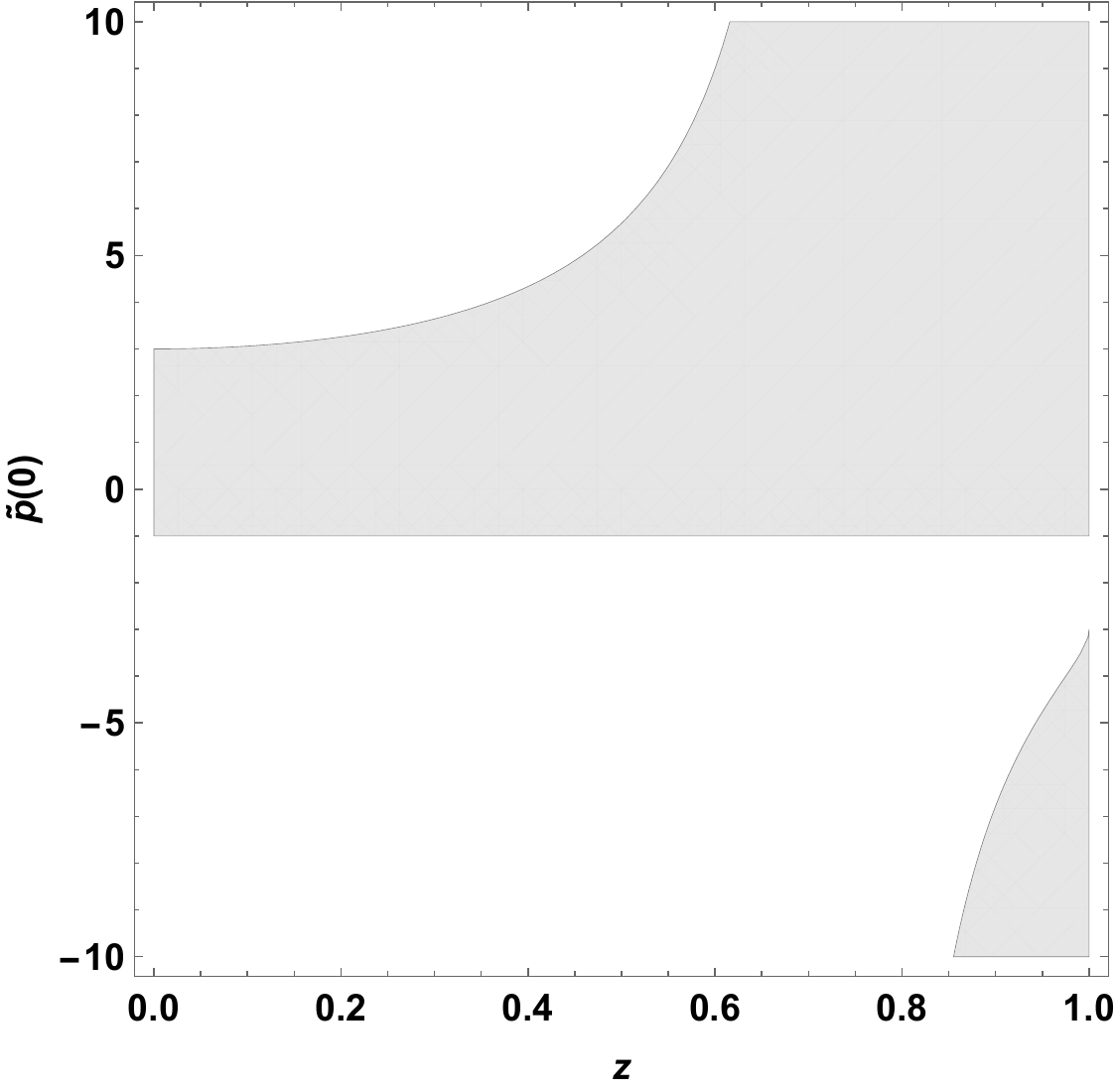}
        %\caption{Potential $A_V$ as a function of $a$.}
    \caption{For the regular black hole, the strong and dominant energy condition is satisfied for the shaded region, i.e. for a wide class of regular initial condition $\Tilde{p}(0)$. Here, the horizontal axis depicts the newly defined radial coordinate $z$, and the vertical axis depicts the initial condition $\Tilde{p}(0)$. $z_b=1/\sqrt{\Tilde{F}_0}$, and $\Tilde{F}_0=1$.}
    
    \label{rfidtag_testing}
\end{figure}
%%%%%%%%%%%%%%%%%%%%%%%%%%%%%%%%%%%%%%%%%%

Let $S \subset \Sigma \subset \mathcal{M}_1$ be an embedded two-dimensional spacelike submanifold. 
$\forall$ $p\in S$, $\exists$ precisely two future-directed null vectors $l$ and $n$ $\in T_p\mathcal{M}_1$  orthogonal to $S$, i.e. $\Tilde{g}_{1~p}(l,\kappa)=\Tilde{g}_{1~p}(n,\kappa)=0$ $\forall$ $\kappa\in T_p S$, along with $\Tilde{g}_{1~p}(l,l)=\Tilde{g}_{1~p}(n,n)=0$, and $\Tilde{g}_{1~p}(l,n)=-1$. Additionally, one of these null vectors, say $l$, satisfies $\Tilde{g}_{1~p}(N,l)\geq 0$. Here $N\in T_p\Sigma$ is normal to $S$ in $\Sigma$ that points outwards from $S$. We now make a continuous assignment of vectors such as $l$ and $n$ $\in T_p\mathcal{M}$ overall points $p\in S$, obtain two distinct vector fields $v_1:S \subset\mathcal{M}_1\to T\mathcal{M}_1$ and $v_2:S\subset \mathcal{M}_1\to T\mathcal{M}_1$, and thereby define two distinct families of null geodesics (null geodesic congruence) that we refer to as outgoing (the one generated by null vectors of the kind $l$) and ingoing families respectively. We say that $S$ is an MTS if it is compact and the expansion of both the null geodesic congruence is non-positive.   

We have the equilibrium configuration $\mathcal{E}=\left(\mathcal{M}_1,\Tilde{g}_1\right)$, where $\Tilde{g}_1$ is given by Eq.(\ref{st2}). Consider the set $\mathcal{M}_1=\mathcal{B}\times \mathbb{R}$, where
$\mathcal{B}\subset \Sigma$ ( $\Sigma$ is the Cauchy surface), is compact and defined as
\begin{equation}
    \mathcal{B}=\left\{q(z,\theta, \phi)\in \Sigma \hspace{0.1cm} \vert \hspace{0.1cm} z\leq z_b\right\}.
\end{equation}

Then, the expansion of the outgoing radial null geodesic congruence at points in $\mathcal{M}_1$ (identified only by the coordinate $z$) is given by
\begin{equation}\label{expansion}
    \Theta(z)=\frac{\sqrt{2}}{z}\sqrt{1-\frac{\Tilde{F}(z)}{z}}.
\end{equation}
 $\Theta$ does not depend on other coordinates due to the symmetry: the existence of the Killing vector fields $\partial_t$, $\partial_{\theta}$ and $\partial_{\phi}$. For a regular end-state equilibrium configuration obtained through the gravitational collapse of a perfect fluid, trapped surfaces (constant $z$ surfaces for which $\Theta(z)<0$) do not form, as seen from the Eq.(\ref{expansion}). Only MTS's (constant $z$ surfaces for which $\Theta(z)=0$) can form corresponding to $z$ that are roots of 
 \begin{equation}
 V(z)=z-\Tilde{F}(z).    
 \end{equation}
 Eq.(\ref{expansion}) tells us that $\Theta(z)$ can take the following values:
\begin{equation}
    \Theta(z)=0, \hspace{0.2cm} \Theta(z)\in \mathbb{R}^+, \hspace{0.2cm} \textrm{or} \hspace{0.2cm} \Theta(z)=\infty.
\end{equation}
 For $\mathcal{O}(\Tilde{F}) \sim \mathcal{O}(z^3)$, which is a necessary and sufficient criterion for maintaining the boundedness of the density throughout the spacetime, i.e. $\rho(z)<\infty$ $\forall$ $z\in [0, z_b]$, we have from Eq.(\ref{expansion}) that $\lim_{z\to 0}\Theta(z)=\infty$. Equivalently, for $\lim_{z\to 0}\Theta(z)<\infty$, one should have $\mathcal{O}(\Tilde{F}) < \mathcal{O}(z^3)$, subsequently implying that $\lim_{z\to 0}\rho(z)=\infty$.
 
An example of a  (bounded density and formation of MTS) can now be constructed as an end-state of the gravitational collapse of a perfect fluid. 
Let $\Tilde{F}(z)=\Tilde{F}_0 z^3$, $\Tilde{F}_0>0$, and $z_b=1/\sqrt{\Tilde{F}_0}$. This choice corresponds to a constant density throughout $\mathcal{M}_1$ previously obtained in 
\cite{Malafarina_2015}. The interior static spacetime $(\mathcal{M}_1,\Tilde{g}_1)$ is glued to the exterior generalized Vaidya spacetime $(\mathcal{M}_2,\Tilde{g}_2)$ such that the Israel-Darmois conditions are satisfied at the timelike hypersurface (denoted by $\Tilde{\Xi}$ with $\Tilde{\Xi}\hookrightarrow\mathcal{M}_1$ as well as $\Tilde{\Xi}\hookrightarrow\mathcal{M}_2$) and defined by $z=z_b$. These conditions impose the following restrictions:
\begin{equation}\label{mc21}
    \mathcal{R}=z,
\end{equation}
\begin{equation}\label{mc22}
    \Tilde{F}(z)=2\Tilde{M}_V(z,v),
\end{equation}
and
\begin{equation}\label{mc23}
    \Tilde{M}_V(z,v),_z=\frac{\Tilde{F}(z)}{2 z}+z e^{2\Tilde{\nu}(z)}\left(\frac{d\Tilde{\nu}(z)}{dz}\right) \sqrt{\Tilde{G}(z)}.
\end{equation}
Here $\Tilde{M}_V(z,v)$ is such that
\begin{equation*}
    M_V(\mathcal{R},v)\to \Tilde{M}_{V}(z,v),
\end{equation*}
as $\mathcal{X}(t,r)\to \mathcal{X}_e(r)=\Tilde{\mathcal{X}}(z)$ in (\ref{tendto}). The generalized Vaidya mass corresponding to $\Tilde{g}_2$ is then obtained by integrating Eq. (\ref{mc23}) along with using Eqs. (\ref{mc21}) and (\ref{mc22}). The total spacetime is then $(\mathcal{M},\Tilde{g})$, where $\mathcal{M}=\mathcal{M}_1\sqcup \mathcal{M}_2$. Also 
$\Tilde{g}\equiv \Tilde{g}_1$ $\forall ~p\in \mathcal{M}$ with $z\leq z_b$ and $\Tilde{g}\equiv \Tilde{g}_2$ $\forall ~p\in \mathcal{M}$ with $\mathcal{R}\geq z_b$.

$(\mathcal{M}, \Tilde{g})$ then has the following properties:

\begin{enumerate}
    \item $\Theta(z_b)=0$. Hence, $z=z_b$ corresponds to MTS. $\Hat{\mathcal{M}}_1=\Tilde{\mathcal{M}}_1\setminus\Tilde{\Xi}$ does not intersect the causal past of any point $p\in \Tilde{\mathcal{M}}_2\setminus \Tilde{\Xi}$ because of the existence of MTS.
    
    \item The density ($\Tilde{\rho}=3\Tilde{F}_0$) is positive and finite throughout $\mathcal{M}_1$ as seen from Eq.(\ref{eq1}). The choice of $F(z)$ (following (\ref{orderF}) that gives rise to such density is what leads to non-admittance of the singularity by $\mathcal{M}_1$.
    
    \item The pressure profile of the interior spacetime is obtained by solving the differential Eq.(\ref{dep}) with the initial condition $\Tilde{p}(0)=F_0\frac{1-3k}{k-1}$ as
    \begin{equation}\label{pp}
        \Tilde{p}(z)=-\frac{F_0\left(1-3k\sqrt{1-F_0 z^2}\right)}{1-k\sqrt{1-F_0 z^2}}.
    \end{equation}
    Here, we have the freedom to choose the initial data in terms of the initial pressure $\Tilde{p}(0)$ or in terms of $F_0$ and $k$. The equation of state of the equilibrium matter field for various choices of $\Tilde{p}(0)$ is depicted in Fig.(1). The strong and the dominant energy conditions for such matter field are satisfied for a wide range of this initial data $\Tilde{p}(z=0)$ throughout $\mathcal{M}_1$, as shown in Fig.(2). This depicts that we do not require an exotic matter field to obtain $\mathcal{M}_1$. In other words, violation of the energy conditions (strong, weak, dominant) is not necessary to obtain $\mathcal{M}_1$. However, it is worth mentioning that negative pressures come into the picture, as seen in Fig.(1).
    
    \item The curvature scalars does not blow up $\forall$ $p\in \mathcal{M}_1$ as depicted by the expression of the Ricci scalar:
    \begin{equation}
        R=6F_0\frac{\left(1-2k \sqrt{1-F_0 z^2}\right)}{1-k \sqrt{1-F_0 z^2}}.
    \end{equation}

    \item $(\mathcal{M}_1, \Tilde{g}_1)$ (and hence $(\mathcal{M},\Tilde{g})$) is geodesically complete. To depict this, we follow the procedure by Zhou and Modesto
    \cite{Modesto_2022} 
    (Refer to the Appendix).

% \item The collapsing perfect fluid spacetime Eq.(\ref{st1}) can be matched smoothly with the exterior generalized Vaidya spacetime
%    \cite{Wang_1999}
%    such that the union forms a valid solution to Einstein's field equations
%    \cite{Joshi_1999, Goswami_2007}.
%By smooth matching, we mean the satisfaction of the Darmios-Israel junction conditions
%\cite{Darmios_27, Israel_67}. 
%These conditions impose restrictions on the generalized Vaidya mass function. (Please refer to  \cite{Joshi_1999, Goswami_2007} for further information on the matching and the restrictions on the generalized Vaidya mass function.)

 \item A null hypersurface $\mathcal{K}\hookrightarrow\mathcal{M}_1$ is called a Killing horizon if $\exists$ a Killing vector field $\chi\in \Gamma\left(T\mathcal{M}_1\right)$ such that its norm with respect to $\Tilde{g}_1$ is zero $\forall$ $p\in \mathcal{K}$. In a static spacetime with metric $\Tilde{g}_1$, if $\exists$ an event horizon, then it is a Killing horizon corresponding to the Killing vector field $\partial_t$
    \cite{Carter}. 
    Using Eq.(\ref{nu}), we obtain exactly one Killing horizon
    $$\mathcal{K}:\{p\in\mathcal{M}_1\hspace{5pt}\vert \hspace{5pt} \Tilde{g}_{1~p}(\partial_t,\partial_t)=0 \},
    $$
    defined by 
    \begin{equation}
        z=\frac{1}{\sqrt{F_0}}\left(1-\frac{1}{k^2}\right)^{1/2}.
    \end{equation}
    Hence, $z=z_b=1/\sqrt{F_0}$ is not the Killing Horizon and, therefore, not the event horizon. %Note that the question of whether $\mathcal{M}$ admits an event horizon is relevant only if it is asymptotically flat. Here, we have generalized Vaidya spacetime in the exterior, which may or may not be asymptotically flat, depending on the choice of the free function (generalized Vaidya mass function) $M(\mathcal{R},v)$ constrained by a partial differential equation and two boundary conditions as shown in 
  %  \cite{Joshi_1999, Goswami_2007}.

    \item $(\mathcal{M}, \Tilde{g})$ is obtained as a limiting case of $(\mathcal{M}, g)$, or in other words, such spacetime is obtained as an end state of a class of regular initial data. One such element of this class is the canonical configuration identified by a fixed radial profile $p(t,r)=p(r)=\Tilde{p}(z=R_e(r))$ (similar to the choice as shown in 
    \cite{Malafarina_2015}).

    \item Such RBH is equivalent to a little modification to the Schwarszchild interior solution (extending the largest radial coordinate $z_b$) that ultimately gives rise to the existence of an MTS (and hence an RBH). Of course, RBH has been studied before, but none has been shown to arise from gravitational collapse. 
 %   \textcolor{black}{The construction in the present paper complements the construction by Simpson and Visser 
%    \cite{Simpson}, 
%    in which the $r$ Schwarzschild coordinate in Schwarzschild spacetime is replaced with $\sqrt{r^2+l^2}$ to get rid of the singularity.} 
\end{enumerate}
\section{Outlook}
 In conclusion, we have shown that we can obtain a regular black hole spacetime $(\mathcal{M},\Tilde{g})$ (that is characterized by 1) nonexistence of any future-directed causal curve $\gamma$ from arbitrary $p\in \mathcal{M}_1$ to $\mathcal{M}_2-\Xi$ due to the existence of MTS at $z=z_b$, and 2) nonexistence of incomplete geodesics) as an end state of the dynamical evolution of a perfect fluid. We note that, in our view, such a scenario is more likely to arise in the formation of supermassive black holes at the galactic centres rather than the catastrophic collapse of massive stars at the end of their life cycles. These latter objects would typically end in the formation of spacetime singularities, either covered or naked. %We plan to discuss this separately elsewhere.

 \section{Appendix}
Here we show that $(\mathcal{M}_1, \Tilde{g}_1)$ (and hence $(\mathcal{M},\Tilde{g})$) is geodesically complete. Let us consider a radially infalling null geodesic $\gamma\subset \mathcal{M}_1$. In terms of the conserved quantity $E=-\Tilde{g}_1(\partial_t, u)$ associated with the Killing vector field $\partial_t$ (where $u$ is the tangent vector field along $\gamma$), the equations of motion of such a geodesic is given by
    \begin{equation}\label{rnge}
        \left(\frac{d z}{d\lambda}\right)^2=e^{-2\nu} E^2 \left(1-\frac{\Tilde{F}}{z}\right),
    \end{equation} 
    and
\begin{equation}\label{rnge2}
    \left(\frac{dt}{d\lambda}\right)^2=E^2 e^{-4\nu}.
\end{equation}
    Here, $\lambda$ is the affine parameter along the geodesic. Solving the Eq.(\ref{eq2}) along with using Eq.(\ref{pp}), we obtain an explicit coordinate expression of $\nu$ as 
    \begin{equation}\label{nu}
        \nu(z)=C_0+\ln{\left(1-k\sqrt{1-F_0 z^2}\right)},
    \end{equation}
    where $C_0$ is a constant of integration. Integrating Eq.(\ref{rnge}), we get the relation between the radial coordinate $z$ and the affine parameter $\lambda$ along $\gamma$ as
    \begin{equation}
       \lambda=\frac{e^{C_0}}{E}\left(kz-\frac{\sin^{-1}{\sqrt{F_0 z}}}{\sqrt{F_0}} \right)+\frac{C_1}{E},
    \end{equation}
   where $C_1$ is another constant of integration. The above equation shows that $\gamma(\lambda=C_1/E)$ corresponds to $z=0$. We highlight the following important point: The radially infalling photon reaches the centre $z=0$ in finite proper time. We could have safely said that $\gamma$ is complete if it were infinite proper time. However, we now investigate if extending $\gamma$ beyond $z=0$ is possible for our case. If the answer to this question is affirmative, then we have the following: When $\lambda$ crosses $C_1/E$ from below (i.e. $\lambda< C_1/E$), the coordinates along $\gamma$ make the transition $\theta \to \pi-\theta$ and $\phi \to \pi+\phi$. Additionally, $dz/d\lambda$ flips the polarity from negative to positive. This polarity flip is because for initially radially infalling null geodesic, $z$ decreases with increasing $\lambda$ until $z=0$, then increases with the further increment of $\lambda$. 

   Now, upon Taylor expanding $dt/d\lambda$ around $z=0$, we obtain
   \begin{equation}
       \frac{dt}{d\lambda}=\sum_{i=0}^{\infty} \frac{z^i}{i!}\frac{d^i}{dz^i}\left(\frac{dt}{d\lambda}\right)\Big|_{z=0}.
   \end{equation} 
   Let us assume that $\exists$ at least one odd term
   \begin{equation}
      \frac{z^{2n-1}}{(2n-1)!}\frac{d^{2n-1}}{dz^{2n-1}}\left(\frac{dt}{d\lambda}\right), \hspace{0.5cm} n\in \mathbb{N},
   \end{equation}
   in this Taylor expansion, that does not vanish. Consider
   \begin{equation}
   n_0=    \textrm{min} \{m \hspace{0.2cm} \vert \hspace{0.2cm} \frac{d^{m}}{dz^{m}}\left(\frac{dt}{d\lambda}\right) \neq 0\, \hspace{0.2cm} m= 2n-1, \hspace{0.2cm} n\in \mathbb{N}\}.
   \end{equation} 
   Then 
   \begin{equation}
       \frac{d^{n_0}}{dz^{n_0}}\left(\frac{dt}{d\lambda}\right)
   \end{equation}
can be written as the sum of terms containing lower (even) derivatives of $\frac{dt}{d\lambda}$ and the term 
\begin{equation}
    \left( \frac{dz}{d\lambda}\right)^{n_0}\frac{d^{n_0}}{dz^{n_0}}\left(\frac{dt}{d\lambda}\right).
\end{equation}
E.g., if $n_0=5$, then
    \begin{align}\label{tye}
    \frac{d^{5}}{dz^{5}}\left(\frac{dt}{d\lambda}\right) =& 5\left(\frac{dz}{d\lambda}\frac{d^4z}{d\lambda^4}+2\frac{d^2z}{d\lambda^2}\frac{d^3z}{d\lambda^3}\right) \frac{d^2}{dz^2}\left(\frac{dt}{d\lambda}\right)+10 \left(\frac{dz}{d\lambda}\right)^3\frac{d^2z}{d\lambda^2}\frac{d^4}{dz^4}\left(\frac{dt}{d\lambda}\right)+\\
    &\left(\frac{dz}{d\lambda}\right)^5\frac{d^5}{dz^5}\left(\frac{dt}{d\lambda}\right).
\end{align}
We now have 
\begin{equation}
    \lim_{\lambda \to \lambda_0^+} \frac{d^{5}}{dz^{5}}\left(\frac{dt}{d\lambda}\right) \neq \lim_{\lambda \to \lambda_0^-} \frac{d^{5}}{dz^{5}}\left(\frac{dt}{d\lambda}\right)
\end{equation}
as seen from the last term of the expansion Eq.(\ref{tye}), since $dz/d\lambda$ changes the polarity as $\lambda$ crosses $C_1/E$ (or the curve crosses $z=0$).  In such a scenario, $dt/d\lambda$ is not an analytic function of $z$. This is the case for Hayward's , as shown in 
\cite{Modesto_2022}.

However, in our case, all the odd terms in the Taylor expansion of $dt/d\lambda$ (Eq.(\ref{rnge2})) vanishes at $z=0$. Hence, $dt/d\lambda$ is a smooth function of $z$ in the neighbourhood of $z=0,$ thereby implying that $\gamma$ is a complete geodesic.

\end{document}